\documentclass[a4, 10pt, twocolumn ]{article}

\usepackage{amsmath, graphicx, hyperref, subfig, fullpage, geometry, sectsty,lpic}
\usepackage{authblk}

\sectionfont{\mdseries\Large\bf\sffamily}
\subsectionfont{\mdseries\large\bf\sffamily}
\subsubsectionfont{\mdseries\bf\sffamily}

\begin{document}

\title{\bf \sffamily Escape path complexity and its context dependency in Pacific blue-eyes ({\it Pseudomugil signifer})}

\author[1]{JE Herbert-Read}
\author[2]{AJW Ward}
\author[1]{DJT Sumpter}
\author[1,3]{RP Mann}
\affil[1]{Department of Mathematics, Uppsala University, Uppsala, Sweden}
\affil[2]{School of Biological Sciences, The University of Sydney, Sydney, Australia}
\affil[3]{Professorship of Computational Social Science, ETH Zurich, Zurich, Switzerland}

\date{}

\maketitle
\section*{Abstract}
The escape trajectories animals take following a predatory attack appear to show high degrees of apparent `randomness' - a property that has been described as `protean behaviour'. Here we present a method of quantifying the escape trajectories of individual animals using a path complexity approach. When fish ({\it Pseudomugil signifer}) were attacked either on their own or in groups, we find that an individual's path rapidly increases in entropy (our measure of complexity) following the attack. For individuals on their own, this entropy remains elevated (indicating a more random path) for a sustained period (10 seconds) after the attack, whilst it falls more quickly for individuals in groups. The entropy of the path is context dependent. When attacks towards single fish come from greater distances, a fish's path shows less complexity compared to attacks that come from short range. This context dependency effect did not exist, however, when individuals were in groups. Nor did the path complexity of individuals in groups depend on a fish's local density of neighbours.  We separate out the components of speed and direction changes to determine which of these components contributes to the overall increase in path complexity following an attack.  We found that both speed and direction measures contribute similarly to an individual's path's complexity in absolute terms. Our work highlights the adaptive behavioural tactics that animals use to avoid predators and also provides a novel method for quantifying the escape trajectories of animals.  

\section*{1. Introduction}

Prey have evolved an array of behaviours in order to escape or avoid predatory attacks such as stotting \cite{fitzgibbon1988stotting}, thanatosis \cite{miyatake2004death} and defensive regurgitation \cite{schmidt1990insect}. But when an attack is inevitable or already initiated, the most common defence a prey uses is to flee, thereby attempting to maximise the instantaneous distance between itself and the threat \cite{weihs1984optimal}. These escape responses involve both non-locomotor and locomotor components \cite{domenici2007hypoxia} and for some animals, are initiated when the apparent looming rate (the rate at which an object of a particular size appears to change on an individual's retina) reaches some threshold \cite{santer2012predator, domenici2002visually}. Non-locomotor components of these behaviours include the escape latency and the reaction distance to the threat, whilst locomotor components include the turning and tangential speeds of an escape path \cite{domenici2007hypoxia}. The timings and directions of these escape responses are context dependent \cite{domenici2010context, eaton1991stimulus} and rely on integrating information on the distance and direction of an approaching threat \cite{hemmi2010multi, domenici2011animal}. The flight initiation distances of grasshoppers, ({\it Psinidia fenestralis}) for example, changes under repeated attacks \cite{bateman2013switching}. Further, the initial escape direction animals take (with regards to the direction of attack) can also be highly variable \cite{domenici2011animal, domenici1993escape, eaton1991stimulus}.  Cockroaches, for example, have multiple preferred directions of escape \cite{domenici2008cockroaches}. In other cases, directions of escape may be limited by the locomotory constraints on an animals movements or obstacles  \cite{domenici2011animal, eaton1991stimulus}.  Escape behaviour, therefore, is a classic example of how an animal can rapidly integrate information from its environment to produce an appropriate behavioural response that is constrained by the animal's biomechanics and information processing capabilities.    

One aspect of the escape response that has been difficult to quantify to date, has been termed, `protean' behaviour \cite{driver1988protean, humphries1970protean}. Animals displaying protean behaviour have escape paths that appear to show high degrees of `randomness'. This randomness has been attributed to animals reducing the predictability of their movements in order to avoid predators intercepting them, or to increase the likelihood that a predator abandons the chase \cite{jones2011prey}. But the difficulty in quantifying these paths have made these observations largely anecdotal, making it difficult to compare these responses to varying conditions and contexts. Whilst instantaneous measures such as the escape direction, tangential speeds, acceleration and turning speeds of an individual can all be measured separately \cite{walker2005faster}, how these variables combine to increase the unpredictability of an animal's path over time remains unclear. These instantaneous measures provide key insights into predatory-prey dynamics and are important in determining whether prey survive following single strikes \cite{webb1976effect, fuiman1993development, eaton1977mauthner, eaton1991stimulus}. However, considering some predators actively chase their prey \cite{domenici2014sailfish, neill1974experiments}, if an animal can sustain high levels of path complexity, then this is likely to be a strong determinant of its survival chances.  

Simply increasing path complexity, however, may not be adaptive under all contexts. When an animal is further from the threat, it has more time to implement an escape plan that could involve seeking cover \cite{rahel1988complex} or freezing \cite{quinn2005personality}.  The presence of neighbours can also affect the escape behaviour of individuals.  When solitary herring are attacked, fish escape in directions facing away from the stimulus less often than when they are attacked in groups \cite{domenici1997escape}. Fish on their own also have a higher proportion of shorter response latencies compared to individuals in groups \cite{domenici1997escape} suggesting that individuals in groups have longer to integrate information on the direction and distance of attack before performing an evasive manoeuvre.  In groups, therefore, individuals may have different path complexities compared to when they are on their own. We set out to quantify how an animal integrates direction and speed changes in its escape path following an attack over longer periods than previously analysed.  In particular, we wanted to know how unpredictable an animal's path became following an attack. We also asked whether the unpredictably of an escape path was context dependent, and changed as a function of the distance from the threat, whether individuals were in groups and depending on the local density of neighbours when in groups.  

\section*{2. Methods}
\subsection*{(a) Experimental procedure}
Pacific blue-eyes ({\it Pseudomugil signifer}) were caught in hand nets from Narrabeen Lagoon, New South Wales, Australia (33�43�03 S, 151�16�17 E) and were housed in 150 l aquaria.  These fish are a facultative shoaling species found both on their own and in groups of various size \cite{pusey2004freshwater, herbert2010sensory}. Fish were held for at least two weeks prior to experimentation.They were kept under a 12:12 dark:light photoperiod and were fed flake food {\it ad libitum}. Fish were fed on the evening after trials had been completed.  An annulus arena (760 mm external diameter, 200 mm internal diameter) was filled to a depth of 70 mm with aged and conditioned tap water.  The stimulus, a 6 cm$^2$ piece of opaque black plastic fixed to the end of a white rod, 4 mm in diameter, was angled so that it could be horizontally extended 200 mm into the arena (at a height of 2-3 cm above the water�s surface). A camera (Logitech Pro 9000) placed directly above the centre of the arena filmed the experiments at 15 frames per second.  The arena was lit by fluorescent lamps and was visually isolated. See Movie S1 for a set-up of the experimental arena. 

For each trial, we placed a single fish, or a group of fish, into the arena and waited for 3 minutes to allow the fish to acclimate to the new environment. During these three minutes, the fish began to explore the arena. Following the acclimation time, we extended the stimulus above the surface of the water, which usually caused the fish to initiate an escape response.  We released the stimulus when the fish were at difference distances from it (distance calculated from the farthest part of the stimulus protruding above the water). Films (n = 77 of individual fish) and groups (n = 39) were converted from .wmv format to .avi using DirectShowSource and VirtualDub (v 1.9.2). Trials of individual fish were subsequently tracked using CTrax \cite{branson2009high}. We manually corrected any errors the tracking software had made using the associated Fixerrors GUI in Matlab, giving the raw {\it x,y} co-ordinates of a fish's position at every time step.  For the trials with groups, we tracked the positions of all fish before the threat using automated tracking software (Didson Tracking Software \cite{handegard2008automated}) and manually tracked the three closest fish to the stimulus at the time of attack using manual tracking software (Tracker v4.81). Any fish that did not move at least 1 cm in the second before the attack were removed from analysis.  This was because small tracking artefacts could artificially inflate a fish's entropy if it was not moving.  In total, 4 fish from individual trials and 4 fish from the group trials were removed.  To calculate the density of neighbours surrounding the focal individuals in groups, we determined the average number of neighbours within 5, 10, or 20 cm of the focal individual up to one second before the stimulus entered the arena. 

\subsection*{(b) Path complexity}
We use the measure of path complexity developed by Roberts {\it et al.} \cite{roberts2004ped}. This measures the information necessary to specify a segment of the path composed of a series of recorded positions. This measure effectively measures how predictable or unpredictable intervals of the movement path are, giving an indication of how well a predator could infer the likely future location of the fish from its recent movements. Straight line movement can be described very simply, with a direction and a distance. Conversely, random motion requires much more information to describe. Natural animal motion lies between these extremes \cite{roberts2004ped, guilford2004ped}. Defining path complexity in this information theoretic manner gives a more fundamental measure of the unpredictability of an animal's motion than related measures such as tortuosity \cite{roberts2004ped}.

The complexity of a path segment is derived by considering an embedding matrix, $M$ containing the recorded positions of the animal over a time window, $t, t+1, \ldots, t+n$. For the results reported in this paper we use a time window of half a second, which we find gives the best balance between the temporal precision in fixing the complexity to the path and the degree of noise in the complexity. At 15fps this gives a time window of eight time steps. The $x$ component of the embedding matrix is a specified from the $x$ co-ordinates of the positions as below:
\begin{equation}
M_x = \begin{bmatrix} x_t & x_{t+1} & \ldots & x_{t+n/2} \\ \vdots & \vdots & \ddots & \vdots \\ x_{t+n/2} & x_{t+n/2+1} & \ldots & x_{t+n} \end{bmatrix},
\end{equation}
with $M_y$ specified similarly from the $y$ co-ordinates. The full embedding matrix is then the concatenation of the two:
\begin{equation}
M = \begin{bmatrix} M_x & M_y \end{bmatrix}.
\end{equation}
Before calculating the complexity of the embedding matrix we first subtract the mean for each column, to focus on variation around the mean position within the window, creating a new matrix $M'$. The complexity of the segment, $H$, is taken as the entropy of the distribution of the singular values, taken from a singular value decomposition of $M'$. We measure entropy in bits, which is equivalent to using base 2 for the logarithm.
\begin{equation}
\begin{split}
M' = USV, \ 
s_i = S_{ii}, \ 
\hat{s_i} = s_i / \sum_{i=1}^n s_i
\end{split}
\end{equation}
\begin{equation}
H = -\sum_{i=1}^n \hat{s_i} \log_2 \hat{s_i}
\end{equation}
This measure of complexity is strictly speaking a property of the path segment, rather than an instantaneous value. In this study we take the complexity at time $t$ to be the entropy of the path segment that ends at time $t$, since this is the first point at which the recorded position from $t$ enters the calculation. Therefore this is when we expect to begin seeing the effect of the stimulus. It is important to note that this measure of path complexity is scale, translation and rotation independent \cite{roberts2004ped}, meaning that the absolute mean position, orientation and speed of the fish within the time window will not alter the complexity.

\subsection*{(c) Separation of directional and speed complexity}
To identify whether the complexity of the escape path is determined by variation in direction or speed, or both, we adapted the path complexity measure to isolate these components. To calculate a purely directional measure of complexity we first reconstructed the escape trajectory, retaining the direction of each movement vector but normalising the displacement per time step to a unit size. Similarly, to isolate the speed component we reconstructed the escape path, retaining the size of displacement for each movement vector but standardising all movement directions to lie along a straight line. We then applied the path complexity measure to these reconstructed paths to calculate directional or speed complexity respectively. 

\subsection*{(d) Statistics} 
We used Pearson correlations to assess the relationship between path complexity and the variables we were interested in. Because we performed many of these correlations (n = 18), we reduced the significance level using a Holm-Bonferroni correction to control the likelihood of Type 1 errors, retaining a 0.05 Type 1 threshold.
  
\section*{3. Results}
\begin{figure}
\begin{center}
\subfloat{
                \includegraphics[width=0.45\textwidth]{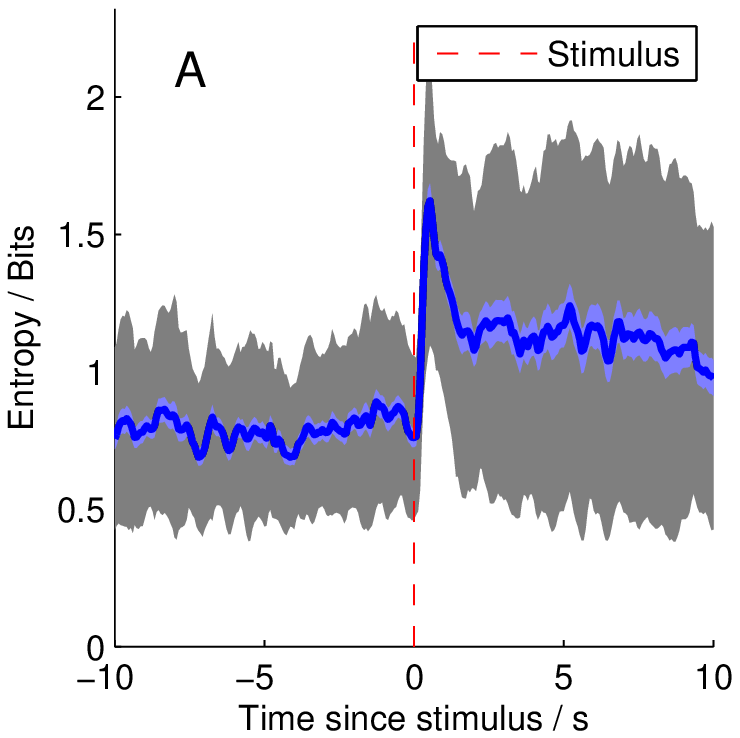}} \\
\subfloat{
                \includegraphics[width=0.45\textwidth]{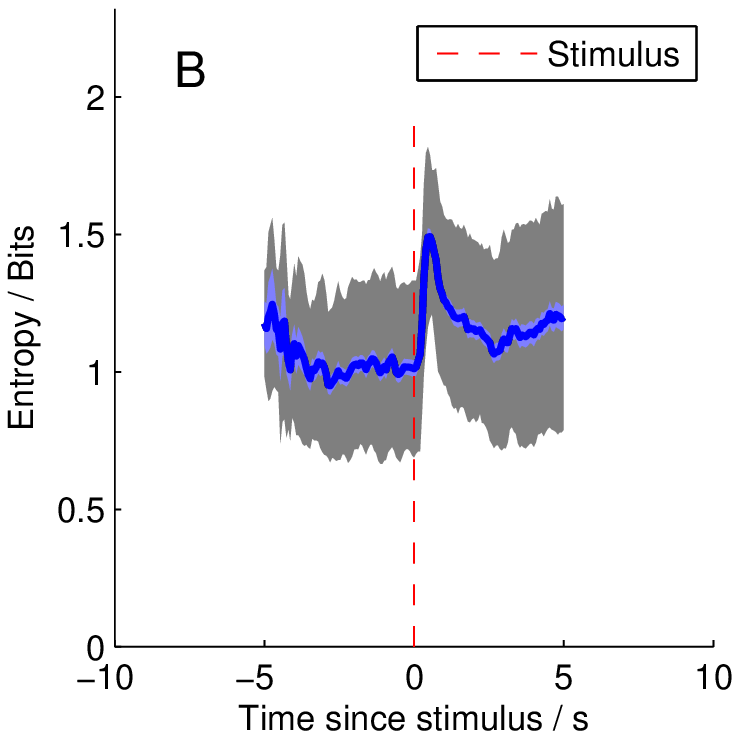}}
\caption{(A) Path complexity for the 73 experiments on single fish, before and after the stimulus, showing mean (blue line), standard error (blue shade) and standard deviation (grey shade). The path complexity rises sharply at the moment of the stimulus as the fish flees, and then remains elevated for at least ten seconds afterwards, showing sustained protean behaviour. (B) Path complexity as in (A) but now for the individuals in group trials (n  = 113 individual fish from 39 group trials). The entropy is generally higher for fish in groups, even before the attack, but again shows a characteristic increase in entropy at the time of attack. In groups, this entropy returns to pre-level attacks more quickly than for individuals on their own.}
\label{fig:stimulus}
\end{center}
\end{figure}

Following the introduction of the stimulus, the fish subsequently induced evasive behaviour (Supp. Video 1). Figure \ref{fig:stimulus}A shows how the path complexity varies with respect to the time of the stimulus, averaged over all individual fish to obtain mean, standard deviation and standard error. A clear peak in complexity occurs directly after the introduction of the stimulus, followed by a sustained period during which the complexity is greater than before the stimulus. This shows that a fish's movements become less predictable in response to the perceived threat and that this new movement pattern is sustained for at least ten seconds after a fish's initial change in direction. Individuals in groups had higher path complexity than individuals on their own even before the stimulus entered the arena, but again showed a clear peak in entropy following the attack (Figure \ref{fig:stimulus}B). In contrast to individuals on their own, the path complexity following the attack returned to pre-stimulus levels more quickly when in groups (Figure \ref{fig:stimulus}B).  

We then asked whether these movement paths were more complex depending on the distance a fish was from the stimulus.  It may be beneficial for a fish that is closer to the threat to increase its path complexity because a small fish can out-manoeuvre a larger predator  \cite{webb1990locomotion,domenici2001scaling}. Conversely, if a fish is further away from the threat, then it may have a better chance of escape by simply fleeing directly away in order to seek cover thereby breaking the line of sight between itself and the threat.  We may therefore expect to see a greater degree of path complexity in fish that were closer to the initial threat, compared to those further away.

\begin{figure}[h!]
\begin{center}
\includegraphics[width=0.45\textwidth]{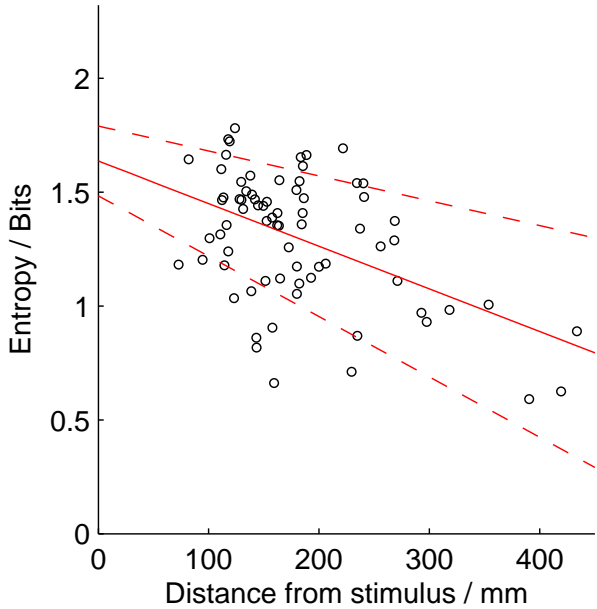}
\caption{Variability in path complexity of single fish immediately after the stimulus with distance from the final position of the stimulus. A significant negative correlation (Pearson's R = -0.49 , P $<$ 0.0001) shows that the fish closest to the threat exhibit the most `random' movement in their escape path.}
\label{fig:proximity}
\end{center}
\end{figure}

Figure \ref{fig:proximity} shows that this prediction is confirmed in the escape responses of fish on their own. Whilst there was no proximity-dependent variation in path complexity prior to the stimulus (Pearson's R = 0.08, n = 73, P = 0.48), there was a negative correlation between the average complexity of the path (for a duration of one second after the stimulus had been released) and the distance from the threat (Pearson's R = -0.49, n = 73, P $<$ 0.0001). Fish that were closer to the threat had more complex paths than those further away from the threat. This effect did not exist, however, for individuals in groups. There was no correlation between the distance to the threat and path complexity for individuals in groups, either before or after the stimulus entered the arena (Before the stimulus: Pearson's R = -0.15, n = 113, P = 0.1; After the stimulus: Pearson's R = -0.01, n = 113, P = 0.91). 

Because the distance to the threat did not affect the path complexity of individuals in groups, we asked instead whether social factors influenced the path complexity of individuals in groups.  In particular, we asked whether the local density of neighbours influenced an individual's path complexity. To test this we correlated a focal fish's average number of neighbours within a 5, 10 or 20 cm radius in the second before the attack, with its average path complexity within the second before or after the stimulus.  For individuals in groups, a fish's path complexity was not correlated with its density of neighbours within any of the distances we chose either before of after the stimulus (Before stimulus, 5 cm radius: Pearson's R = -0.16, n = 113, P = 0.08; Before stimulus 10 cm radius: Pearson's R = -0.16, n = 113, P = 0.08; Before stimulus, 20 cm radius: Pearson's R = -0.17, n = 113, P = 0.08; After stimulus, 5 cm radius: Pearson's R = -0.01, n = 113, P = 0.88; After stimulus, 10 cm radius: Pearson's R = 0.02, n = 113, P = 0.8; After stimulus, 20 cm radius: Pearson's R = -0.02, n = 113, P = 0.85).  

\begin{figure}[h!]
\begin{center}
\subfloat{
                \includegraphics[width=0.45\textwidth]{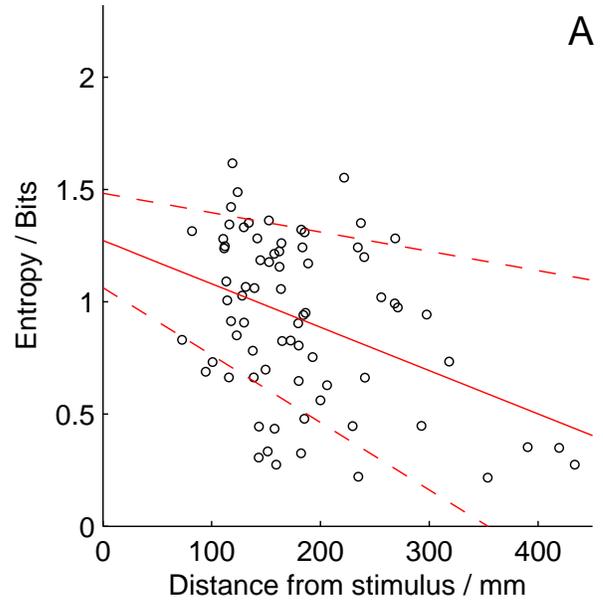}} \\
\subfloat{
                \includegraphics[width=0.45\textwidth]{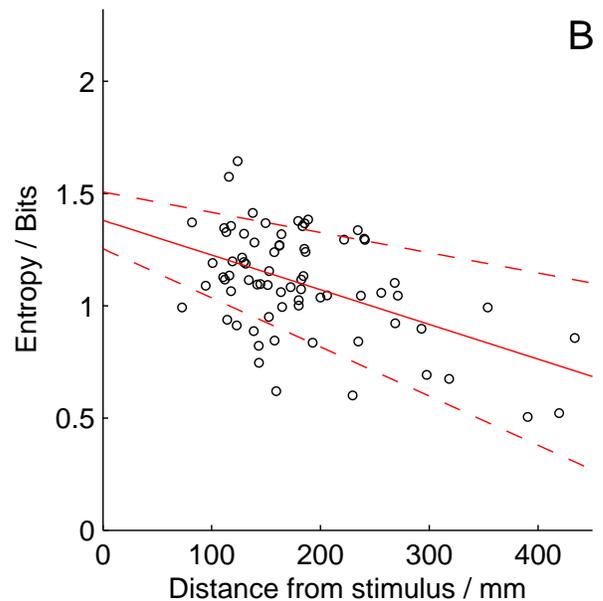}}
             
\caption{Context-dependency in directional (A) and speed (B) complexity after the stimulus, relative to the distance of the fish from the stimulus at the moment of activation. Speed complexity shows a more significant correlation with distance (R = -0.49, n = 73, P = 0.000009) than directional complexity (Pearson's R = -0.39, n =73, P = 0.0006).}
\label{fig:anglespeed}
\end{center}
\end{figure}

\begin{figure}[h!]
\begin{center}
\subfloat{
                \includegraphics[width=0.45\textwidth]{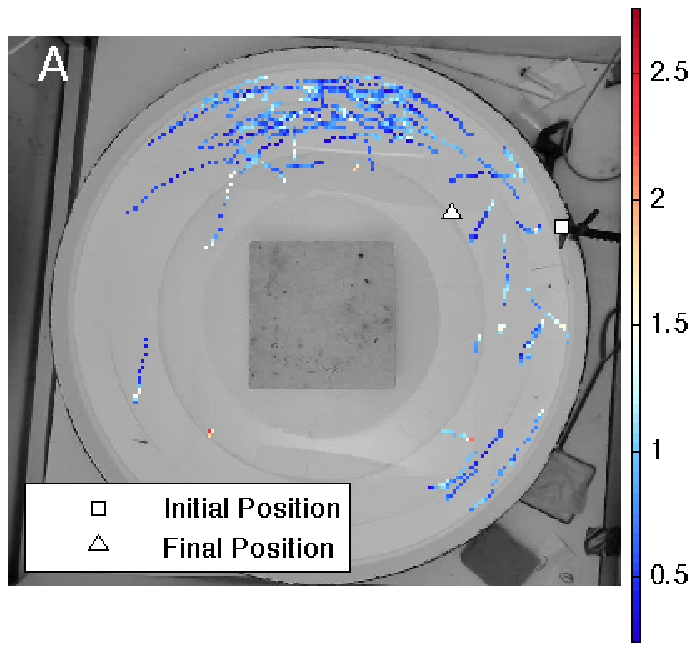}} \\
\subfloat{
                \includegraphics[width=0.45\textwidth]{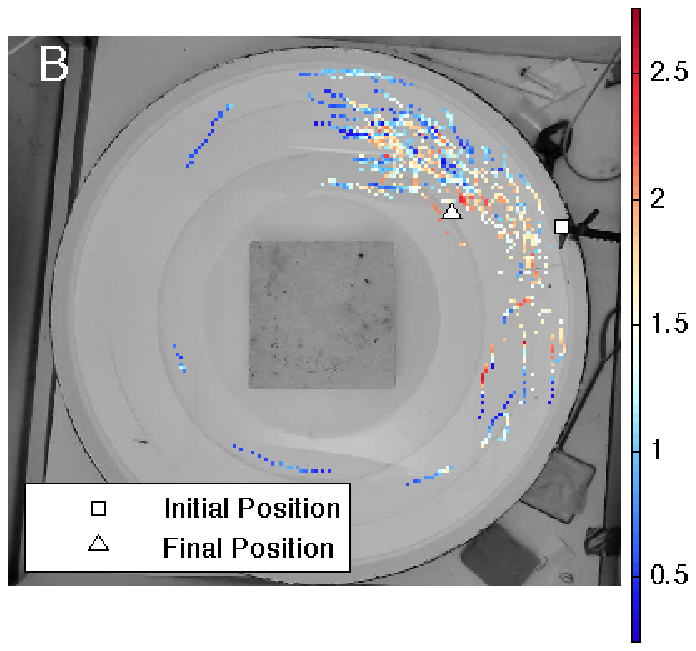}}
             
\caption{The spatial distribution of path complexity before (A) and after (B) the stimulus. The path complexity in a spatial region (5 mm $\times$ 5 mm) is calculated as the mean complexity of the paths passing through that region at the time of passage, over a duration of one second after the initial stimulus. Complexity is substantially higher after the stimulus, and high complexity regions are predominantly clustered near the stimulus' location.}
\label{fig:spatial}
\end{center}
\end{figure}

To assess whether directional or speed variability, or both, was responsible for this context-dependent increase in entropy for individuals on their own, we isolated the directional and speed components of the path complexity by reconstructing the escape paths to exclude speed or directional variation respectively (see Methods) and reran our analysis on these reconstructed escape paths. The results of analysing the reconstructed paths show that for both components, path complexity follows a temporal pattern very similar to Figure \ref{fig:stimulus}, with both components rising quickly at the moment of stimulus and remaining elevated afterwards (Figure S1). Analysis of the context-dependency, as shown in Figure \ref{fig:anglespeed} shows that both variability in speed and directional complexity are significantly associated with the fishes' position relative to the stimulus (directional: Pearson's R = -0.39, n = 73, P = 0.0006; speed: R = -0.49, n = 73, P = 0.000009).  Whilst there was no correlation between the speed complexity and the distance to the stimulus before the attack (Pearson's R = 0.01, n = 73, P = 0.91), after our Holm-Bonferroni correction, the correlation between directional complexity and the distance to the stimulus was marginally non-significant (Pearson's R = 0.34, n = 73, P = 0.0035).  However, we attribute this to one outlying individual that can be observed in Figure \ref{fig:spatial}A in the bottom-left corner of the arena.  

To illustrate both the change in path complexity from before the stimulus to after, and to show the dependence on spatial position relative to the threat, we plotted the average complexity of paths contained within elements of a spatial grid overlaying the experimental arena, both for a duration of one second before the stimulus (Figure \ref{fig:spatial}A) and for one second after the stimulus (Figure \ref{fig:spatial}B). It is clear that, as shown in Figure \ref{fig:stimulus}, the complexity of the paths is substantially higher after the stimulus than before. Moreover, the regions of highest complexity in Figure \ref{fig:spatial}B are those closest to the position of the stimulus.

\section*{4. Discussion}

Following the simulated attack, a fish's path rapidly increased in complexity. For individuals on their own, how much the path increased in randomness was dependent on the fish's distance to the threat; further away from the threat, individuals had lower path complexity, whilst closer to the threat, individuals showed higher levels of randomness. Before the attack, the overall entropy for individuals in groups was generally higher than for individuals on their own. But whilst the path complexity for individuals in groups also increased following the attack, unlike individuals, the escape paths did not vary in entropy depending on their distance from the threat. Nor did a fish's local density of neighbours affect its path complexity either before or after the attack.  For individuals on their own, when partitioning the components of speed and direction, we found that both speed and direction contributed significantly to the overall path complexity during the first second after the onset of the threat. These fish, therefore, can adopt varying escape behaviours to adaptively counteract a range of attack scenarios.  

Both speed and direction contributed to the overall path complexity in the second following the attack. This was evident when we analysed the contribution that speed and direction made towards path complexity depending on the distance an individual was from the threat. Why might it be important to vary both these components? First, it has previously been reported that the range of directions animals take following an attack are limited. This is because predators often attack from the sides of prey \cite{webb1980strike}, making the escape directions usually directed away from the threat \cite{domenici1993escape, domenici1997kinematics}. Second, some animals have multi-modal directions in which they flee, which could initially stop a predator from predicting the initial escape direction \cite{domenici2008cockroaches}, but at the same time, limits the potential number of directions prey can take. Therefore, if a prey were to follow a limited range of directions with the same speed, then a predator would be able to intercept the prey by predicting its position in the future. Relying on direction changes alone, therefore, may not provide enough variation to escape predators that chase their prey.  Changing speed, however, interrupts the predicted interception point, thereby making interception more difficult. Indeed, from a bio-mechanic standpoint, turning can be particularly costly \cite{wilson2013turn} and speed changes may be easier to adapt than direction. Voles ({\it Microtus socialis}) use the same strategy when under attack from barn owls ({\it Tyto alba}) by alternating between freezing and fleeing behaviours, thereby decreasing the predictability of their movements \cite{edut2004protean}.  Since there is a trade-off between speed and the number of direction changes an individual can take \cite{angilletta2008fast}, varying speed may also allow individuals to change direction during times when speed is reduced.  It will now be important to understand how these constraints limit the maximum levels of complexity that selection can achieve.

Before the attack, we found that the complexity of individuals' paths in groups was generally higher compared to when on their own. This is in contrast to the common assumption in many models of collective motion \cite{vicsek1995novel,couzin2002cma,vicsek2012cm}, where individuals reduce the noise in their motion in groups because they follow the motion of others. Here we find that there is higher entropy in the paths of individuals travelling in a group. This can be attributed to individuals having to adjust their position depending on the movements and positions of their neighbours \cite{herbertread2011tro, katz2011its}. Following the attack, the path complexity of individuals in groups increased, but did not show any evidence of being dependent on the distance the individuals were from the threat. After the attack, the entropy returned to pre-attack levels quicker for individuals in groups than individuals on their own. Other aspects of escape behaviour change when individuals are placed into groups. Solitary herring, for example, have shorter response latencies to an attack compared to individuals in groups \cite{domenici1997escape}. Further, the escape directions of individuals in groups are directed away from the threat in 88\% of cases \cite{domenici1994escape}. On their own, however, solitary herring move initially away from the stimulus in only 64\% of cases \cite{domenici1997escape}.  It appears, therefore, that being in a group changes the way information about a threat is perceived and responded to. When individuals are in groups, it may be maladaptive to have substantially increased path complexities for sustained periods, as this may result in individuals becoming separated from the group. Indeed, we observed that complexity levels returned to pre-attack levels more quickly for individuals in groups than for individuals on their own. Future investigations will need to consider how individuals integrate the information on the position and movements of their neighbours \cite{herbert2013role, herbertread2011tro} during these escape responses, and how this alters an individual's path complexity. 

The instantaneous responses of prey to a threat depending on its distance show similar trends to our results.  Webb (1982) classified two behavioural responses, depending on the distance to the threat; either type-I or type-II responses \cite{webb1982avoidance}. Type-I responses were described as behavioural responses that showed relatively slower instantaneous speeds and non-sustained turns.  These responses occurred when predators attacked prey from greater distances \cite{webb1982avoidance}. Type-2 responses, on the other hand, occurred when fast-moving predators attacked from short-range distances which caused the prey to increase their speed and initial turning rate \cite{webb1982avoidance}. These instantaneous measures, therefore, complement our longer time-scale observations of fish's movement paths. Indeed, the path complexity of an individual's movement remains at higher levels to those before the attack for at least 10 seconds demonstrating that the fish are in a heightened state of arousal, perhaps envisioning future attacks. Indeed, multiple attacks are often observed when predators attack prey \cite{handegard2012dynamics}.  Our method highlights the need not only to investigate the initial evasive responses of prey, but also to investigate these behaviours over longer time scales.    

Increasing the complexity of a path following an attack may result in predator-prey arms races becoming tipped in the favour of the prey \cite{dawkins1979arms}. It would be impossible for a predator to predict and intercept a perfectly random path taken by a prey. If prey can perform movement paths that reach these levels of unpredictability, then predators must change their attack mechanism in order to successfully capture their prey. In many cases, this could lead to predators abandoning chasing tactics and instead relying on ambush tactics. Alternatively, predators must evolve new behavioural tactics that improve their manoeuvrability, thereby improving their chance of prey capture \cite{maresh2004high}.  With the method we have used here to quantify the complexity of animal escape paths, we have opened new questions into how prey and predators interact over longer periods of sustained attacks. 

\section*{Acknowledgements}
We thank P. Domenici for useful comments on the manuscript. Code for calculating path complexity was adapted from original work by S.J. Roberts, I. Rezek and R. Freeman. This work was funded by the European Research Council, grant IDCAB 220/104702003 to D.J.T. Sumpter and a University of Sydney starting grant to A.J.W. Ward 
\bibliographystyle{prsb}
\bibliography{blueeyes}

\end{document}